\documentclass[conference]{IEEEtran}
\usepackage{tikz}\usetikzlibrary{shapes,arrows}
\usepackage{epstopdf}
\usepackage{pgfplots}
\usepackage{pgfplotstable}
\usepackage{amsmath}
\usepackage{mathtools}
\usepackage{float}
\usetikzlibrary{decorations.pathreplacing,}
\usepackage{xcolor,colortbl}
\usepackage{tikz}
\usepackage{color}
\usepackage{tikz-3dplot}
\usepackage{fp}
\usepackage{xcolor,colortbl}
\usepackage{caption}
 \usetikzlibrary{arrows, decorations.markings}
\usepackage{amsmath}
\usepackage[linesnumbered,ruled]{algorithm2e}
\usepackage{epstopdf}
\usepackage{epstopdf}
\usepackage{pgfplots}
 \usepackage{amsmath}
 \usepackage{multicol}
\usepackage{caption}
\usepackage{subcaption}
\usepackage{amsmath}
\graphicspath{{Figures/}}
\usetikzlibrary{arrows,automata}
\DeclarePairedDelimiter\floor{\lfloor}{\rfloor}

\usepackage[noadjust]{cite}
\SetCommentSty{mycommfont}
\graphicspath{{Figures/}}%In the preamble
\newcounter{example}[section]
\usepackage{amsmath}
\usepackage{hyperref}

\newcolumntype{M}[1]{>{\centering\arraybackslash}m{#1}}

\begin{document}
 \title{A Novel Beamformed Control Channel Design for LTE with Full Dimension-MIMO }
\author{\IEEEauthorblockN{Pavan Reddy M., Harish Kumar D., Saidhiraj Amuru, Kiran Kuchi}
\IEEEauthorblockA{Department of Electrical Engineering, 
Indian Institute of Technology Hyderabad, India 502285\\
Email:\{ee14resch11005,  ee14mtech11003, asaidhiraj, kkuchi\}@iith.ac.in}
}%
\maketitle
\begin{abstract}
The Full Dimension-MIMO (FD-MIMO) technology is capable of achieving huge improvements in network throughput with simultaneous connectivity of a large number of mobile wireless devices, unmanned aerial vehicles, and the Internet of Things (IoT). In FD-MIMO, with a large number of antennae at the base station and the ability to perform beamforming, the capacity of the physical downlink shared channel (PDSCH) has increased a lot.  However, the current specifications of the \mbox{$3^{rd}$} Generation Partnership Project (3GPP) does not allow the base station to perform beamforming techniques for the physical downlink control channel (PDCCH), and hence, PDCCH has neither the capacity nor the coverage of PDSCH. Therefore, PDCCH capacity will still limit the performance of a network as it dictates the number of users that can be scheduled at a given time instant.  In Release 11, 3GPP introduced enhanced PDCCH (EPDCCH) to increase the PDCCH capacity at the cost of sacrificing the PDSCH resources. The problem of enhancing the PDCCH capacity within the available control channel resources has not been addressed yet in the literature. Hence, in this paper, we propose a novel beamformed PDCCH (BF-PDCCH) design which is aligned to the 3GPP specifications and requires simple software changes at the base station. We rely on the sounding reference signals transmitted in the uplink to decide the best beam for a user and ingeniously schedule the users in PDCCH.  We perform system level simulations to evaluate the performance of the proposed design and show that the proposed BF-PDCCH achieves larger network throughput when compared with the current state of art algorithms, PDCCH and EPDCCH schemes. 

\ \\ \textit{Keywords: Beamforming, blind decoding,  control channel, rate matching, search space. }
\end{abstract}
\IEEEpeerreviewmaketitle 
\section{Introduction} 
Full Dimension-Multi Input Multi Output (FD-MIMO) is a key technology in achieving larger network throughputs by simultaneously connecting a large number of devices. This has been an active topic in the standardisation activities of 3$^{rd}$ Generation Partnership Project (3GPP). In FD-MIMO, a two dimensional antenna array structure is used that helps in beamforming along both elevation and azimuth directions. With this kind of beamforming, an enhanced multi-user MIMO transmission can be done at the base station to achieve a multi-fold enhancement in the network throughput~\cite{Samsung}. %From Release 13, 3GPP specifications support such beamforming~\cite{1536211, 1536212, 1536213}.
From Release 8 to Release 13,  3GPP has continuously evolved its specifications to enhance the multi-user MIMO feature and thus,  enable a large number of users to be supported by the base station~\cite{ComMag,RSwhite}.  In Release 13, 3GPP specifications support both the azimuth and elevation beamforming for the data channel. Based on a newly introduced channel state information-reference signals (CSI-RS) and the demodulation reference signals (DMRS), the beamforming is performed for the data channel.  With this kind of beamforming,  2-3.6 times gain in the cell throughput is achieved~\cite{Samsung}.

%CSI-RS helps in identifying the best beam for a user and DMRS helps the user to estimate the channel for the data transmitted in that best beam. With this kind of beamforming, 2-3.6 times the capacity of the data channels can be achieved~\cite{Samsung}. 

\begin{figure}[t]
\centering
\begin{subfigure}{0.55\textwidth}
\centering
\includegraphics[height= 3.5cm]{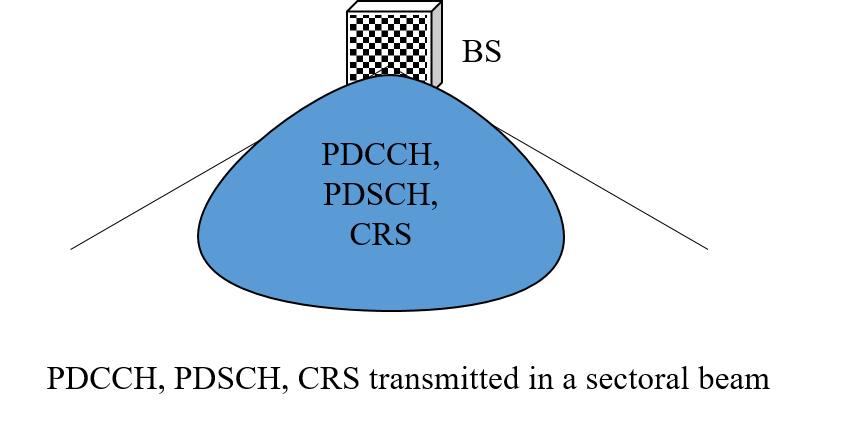}
\caption{Release 8}
\vspace{0.5cm}
\label{fig:EPDCCH}
\end{subfigure}
\begin{subfigure}{0.55\textwidth}
%\centering
\includegraphics[width= 9.2cm]{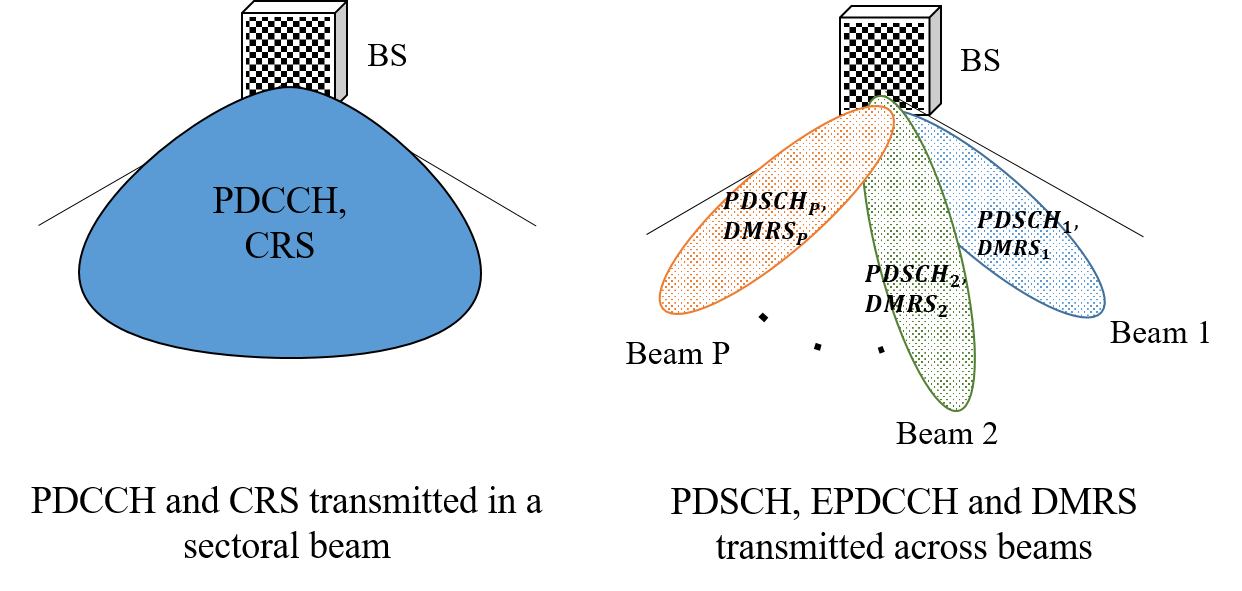}
\caption{Release 13}
\label{fig:EPDCCH}
\end{subfigure}
\caption{Physical layer transmission as per Release 8 and Release 13 3GPP specifications}
\label{fig:Rel813}
\end{figure}
\begin{figure*}[t!]
\centering
\includegraphics[height=3.9cm]{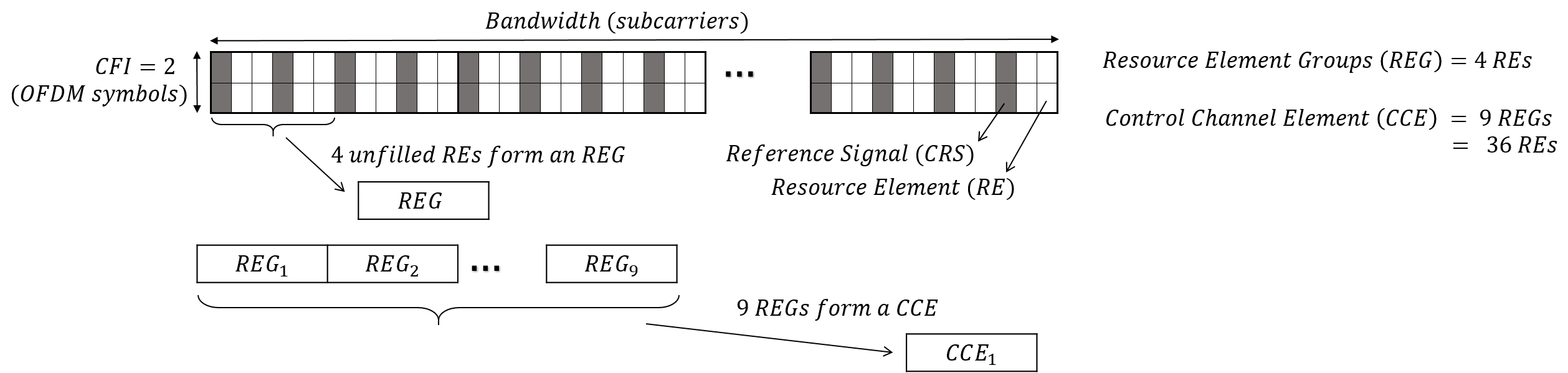}
\caption{ CCE formation in the current LTE PDCCH }
\label{fig:CCE}
\end{figure*}
In Long Term Evolution (LTE), the downlink physical layer has five channels~\cite{1536211, 1536212, 1536213}. They are physical broadcast channel (PBCH) for broadcasting the system information, physical control format indicator channel (PCFICH) for defining the structure of the control channel, physical HARQ indicator channel (PHICH) for conveying the ack/nack, physical downlink control channel (PDCCH) for carrying the control information and physical downlink shared channel (PDSCH) for transmitting the user intended data.  In this paper, we focus on PDCCH which carries the downlink control information (DCI). DCI conveys the information required to decode the user intended data.  As explained later in Section~\ref{sec:3GPP}, the PDCCH region in any subframe is limited to 3 symbols~\cite{1536213} and hence, can accommodate a limited number of DCIs in a transmission time interval (TTI). Thus, the PDCCH effectively indicates the number of users scheduled in any TTI. 
%and thus DCI is crucial in establishing a communication link between the base station and a user. 

In Release 8 3GPP specifications, PDCCH and PDSCH rely on cell-specific reference signals (CRS) for the channel estimation. 
Whereas from Release 13, the PDSCH  supports beamforming and hence, has DMRS for the channel estimation. 
Fig.~\ref{fig:Rel813} presents the transmission of the physical layer signals in both Release 8 and 13. CRS is common for all the users and beamforming CRS would impact the performance of cell search and synchronisation. Thus, with the current 3GPP specifications, the control channel does not possess the benefits of beamforming.  Note that a user can decode the data channel only after decoding a DCI. Thus, even though the beamforming allows to schedule more users in PDSCH, the PDCCH has a limited capacity and has become a bottleneck in increasing the network throughput. In Release 11, to enhance the PDCCH capacity, 3GPP  introduced enhanced PDCCH (EPDCCH) design which uses the concepts of beamforming.  However, the EPDCCH has to be transmitted in the resources of the data channel as shown in Fig.~\ref{fig:EPDCCH}, and the location of the EPDCCH has to be conveyed apriori to the user.

\begin{figure}[t]
\centering
\includegraphics[height= 5.5cm]{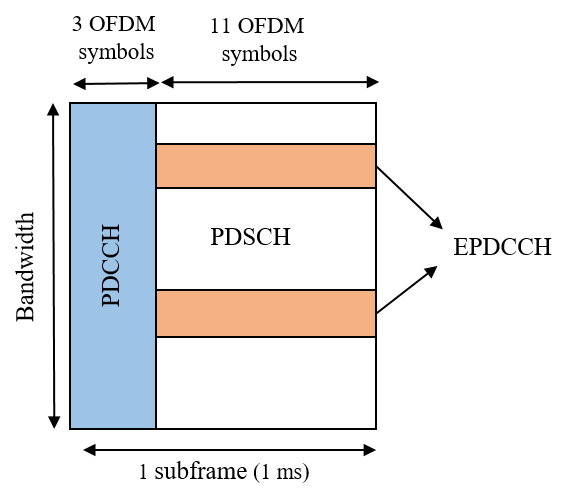}
\caption{EPDCCH configuration in a subframe}
\label{fig:EPDCCH}
\end{figure}

 The availability of the large antennae structure with the FD-MIMO is never exploited in the context of the PDCCH. This is because, for beamforming, some feedback is required from the user. But control channel itself is the first communication link where the user performs blind decoding for the DCI. Improving the PDCCH capacity by exploiting the large antennae structure has a high impact on network throughput and has never been considered in the literature. Motivated by this, in this paper we propose a novel beamformed PDCCH (BF-PDCCH) design which addresses all the above-said issues. The proposed BF-PDCCH design relies on the uplink sounding reference signals transmitted from the user. As explained later in Section~\ref{sec:BF-PDCCH}, the multi-user MIMO feature of proposed BF-PDCCH is enabled only to the users who have sent these uplink reference signals.  The proposed BF-PDCCH design schedules all the other users in a legacy LTE fashion. The features of the proposed BF-PDCCH design are as follows:
\begin{itemize}
\item The proposed BF-PDCCH design is aligned with the 3GPP specifications and requires no changes at the user end.
\item Unlike EPDCCH, the proposed design does not use the resources from the data channel.
\item The proposed BF-PDCCH design significantly increases the control channel capacity.
\end{itemize}

The rest of the paper is organised as follows. Section~\ref{sec:RW} presents some of the related work in the literature. The current physical downlink control channel structure and the enhancements as per 3GPP specifications are explained in the Section~\ref{sec:3GPP}. In Section~\ref{sec:Des}, the antenna array structure, the design and the implementation of the beam weights are explained. In Section~\ref{sec:BF-PDCCH}, the proposed BF-PDCCH design is presented and it's performance is analysed. In Section~\ref{sec:implementation}, the procedures and the algorithms for the implementation of the proposed scheme are discussed.  In Section~\ref{sec:SLS}, the simulation model is presented, and the numerical results are discussed. Some concluding remarks and possible future work are presented in Section~\ref{sec:CFW}.
\section{Related Work}
\label{sec:RW}
In~\cite{Balamurali}, authors have presented an algorithm to optimally schedule the users in PDCCH and thus, increase the control channel capacity.  In \cite{orthPDCCH}, authors have proposed a novel method of allocation for cell radio network temporary identifiers and increase the control channel capacity. In \cite{powerPDCCH}, authors propose power allocation techniques to improve the control channel capacity. However, none of the above papers discussed the beamforming and exploited the large antennae structure for increasing the control channel capacity. Further, we have implemented the optimal LTE-PDCCH scheduling algorithm presented in~\cite{Balamurali}, and compared its performance against the proposed BF-PDCCH design in Section~\ref{sec:SLS}.

In~\cite{EPDCCH3,EPDCCH4}, novel search space designs of EPDDCH have been presented to improve the capacity of the control channel. 
In~\cite{EPDCCH, EPDCCH1}, the performance and analysis of the EPDCCH design is presented. 
In~\cite{EPDCCH2}, the authors have proposed an algorithm to improve the channel estimation accuracy and thus, in turn, improve the performance of the EPDCCH. 
However, as per the 3GPP specifcations~\cite{1536211,1536212,1536213}, the EPDCCH uses the resource elements from the data channel for beamforming. 
To the best of our knowledge, none of the papers in the literature have addressed the issue of increasing the control channel capacity by exploiting the large antennae structure. Next, we explain the current control channel design and it's enhancements  as per 3GPP specifications.

%\cite{EPDCCH1,EPDCCH2,EPDCCH3,EPDCCH4}
%\cite{orthPDCCH}\cite{powerPDCCH}.% \cite{newDCI,newMDCI,unifiedAL}

\section{3GPP Physical Downlink Control Channel}
\label{sec:3GPP}
DCI is the payload transmitted in PDCCH. DCI carries the information for decoding the user data, location of uplink scheduling, random access responses, modulation and coding scheme.  There are various DCI formats for each purpose. Prior to transmission, the DCI payload is appended with cyclic redundancy check parity bits, convolution coded and is then rate-matched to a certain number of bits called aggregation level (AL). These rate-matched bits are then QPSK modulated and multiplexed in the radio frame. 

The PDCCH is present in the first few orthogonal frequency division multiplexing (OFDM) symbols of every subframe. The number of symbols for PDCCH is defined by  PCFICH.  The first OFDM symbol has PCFICH, PHICH and PDCCH multiplexed in it.   In LTE, the smallest time-frequency resource in a radio frame is called as a resource element. Excluding the PCFICH and PHICH resource elements, the remaining resource elements available are grouped in number of four (in a frequency first and time second manner) and called as Resource Element Groups (REGs).  A collection of nine such REGs is called as one control channel element (CCE). In LTE PDCCH, the allocation of the DCIs is done in units of CCEs. Fig.~\ref{fig:CCE} presents the formation of CCEs as per current 3GPP specifications~\cite{1536211}. Based on the channel conditions of the user, the payload is rate matched to an aggregation level. The data in one AL can fit in one CCE. In LTE PDCCH, the AL$\in$\{1,2,4,8\} and thus, AL=2 requires 2 CCEs and so on.  

\begin{figure}[t!]
\centering
\includegraphics[width=4cm]{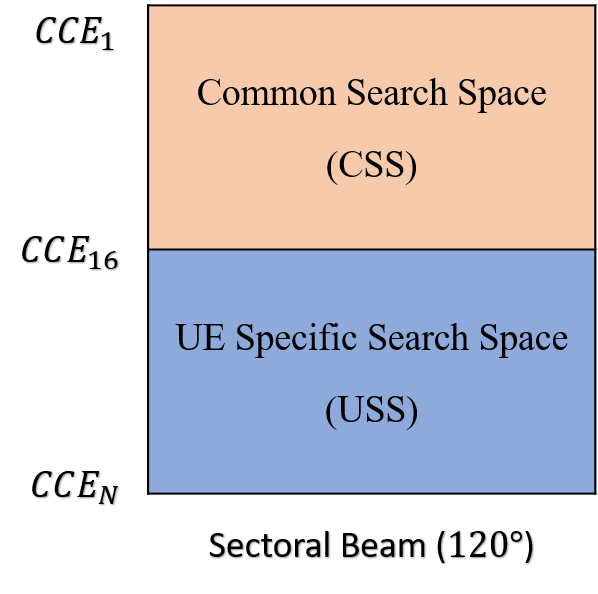}

\caption{ Search space design of current 3GPP PDCCH}
\label{fig:currPDCCH}
\end{figure}
 \begin{figure}[ht]
\centering
\includegraphics[width=9cm]{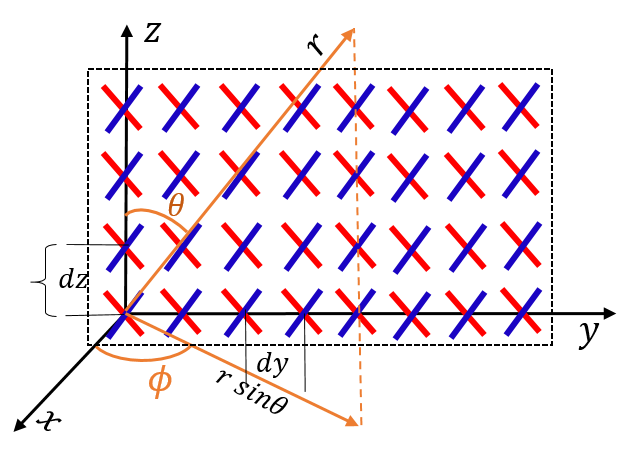}
\caption{Antenna array structure}
\label{fig:AntennaStructure}
\end{figure}
 Search space is defined as the region or collection of CCEs over which a user expects a DCI. The control channel region is broadly classified into two categories. 
 \begin{itemize}
 \item Common Search Space (CSS): Common DCIs which carry system information, paging and common scheduling information are transmitted in a specific region called CSS. 
 \item UE specific Search Space (USS): The DCIs which are intended for a particular user are transmitted in USS. 
 \end{itemize}
 Note that, the number of CCEs in PDCCH varies with the number of symbols over which PDCCH is present and also with the bandwidth of the system. In any subframe, irrespective of the number of CCEs available, the CSS is present only in the first 16 CCEs. The search space design as per the current 3GPP specifications is depicted in Fig.~\ref{fig:currPDCCH}. 

In any search space, the scheduling of DCIs is carried out as follows.  For each DCI, based on the AL and the user identity~\cite{1536211}, possible CCEs for scheduling are calculated using the formula given in (\ref{eqn:bd}). 
The base station can transmit the DCI in any of those possible CCE locations.
\begin{equation}
CCE_{index}=L\{(Y_k+m)mod\floor{N_{CCE,k}/L}\}+i
\label{eqn:bd}
\end{equation}
where, 
$i=0\ldots L-1$, $ L$ is aggregation\  level, $N_{CCE,k} $ is number\  of\  CCEs \ in\  the\  subframe $k$, $Y_k$ and  $m\ $  are\  the\ constants\ defined\ by the higher layer parameters in 3GPP specification~\cite{1536331}.
 
 Note that the user has no information about the location and the aggregation level of the DCI. Hence, the user calculates all the possible indices and blindly performs the search at all those locations. This procedure is repeated for all the aggregation levels until a DCI is decoded.

In any subframe, based on the available bandwidth, there are a limited number of CCEs in PDCCH. This limitation has an impact on the multi-user scheduling in PDSCH.  In Release~11, 3GPP has introduced enhanced PDCCH design to increase the PDCCH capacity. The EPDCCH is transmitted in the data channel region as shown in the Fig.~\ref{fig:EPDCCH}. The search space region for monitoring the DCI in EPDCCH is conveyed to the user prior through higher layer signalling. The advantage of the EPDCCH is that it can use beamforming concepts like data channel and thus, schedule more number of DCIs. However, this comes at the cost of sacrificing the PDSCH resources. Note that the EPDCCH has demodulation reference signals to decode the beamformed data. In this paper, we propose a novel BF-PDCCH design which does not use any PDSCH resources and yet achieves improvement in network capacity. Next, we present the design and implementation of beams for the proposed BF-PDCCH scheme.

\section{Design and Implementation of Beams for BF-PDCCH}
\label{sec:Des}
\begin{figure}[t!]
\centering
\includegraphics[width=6cm]{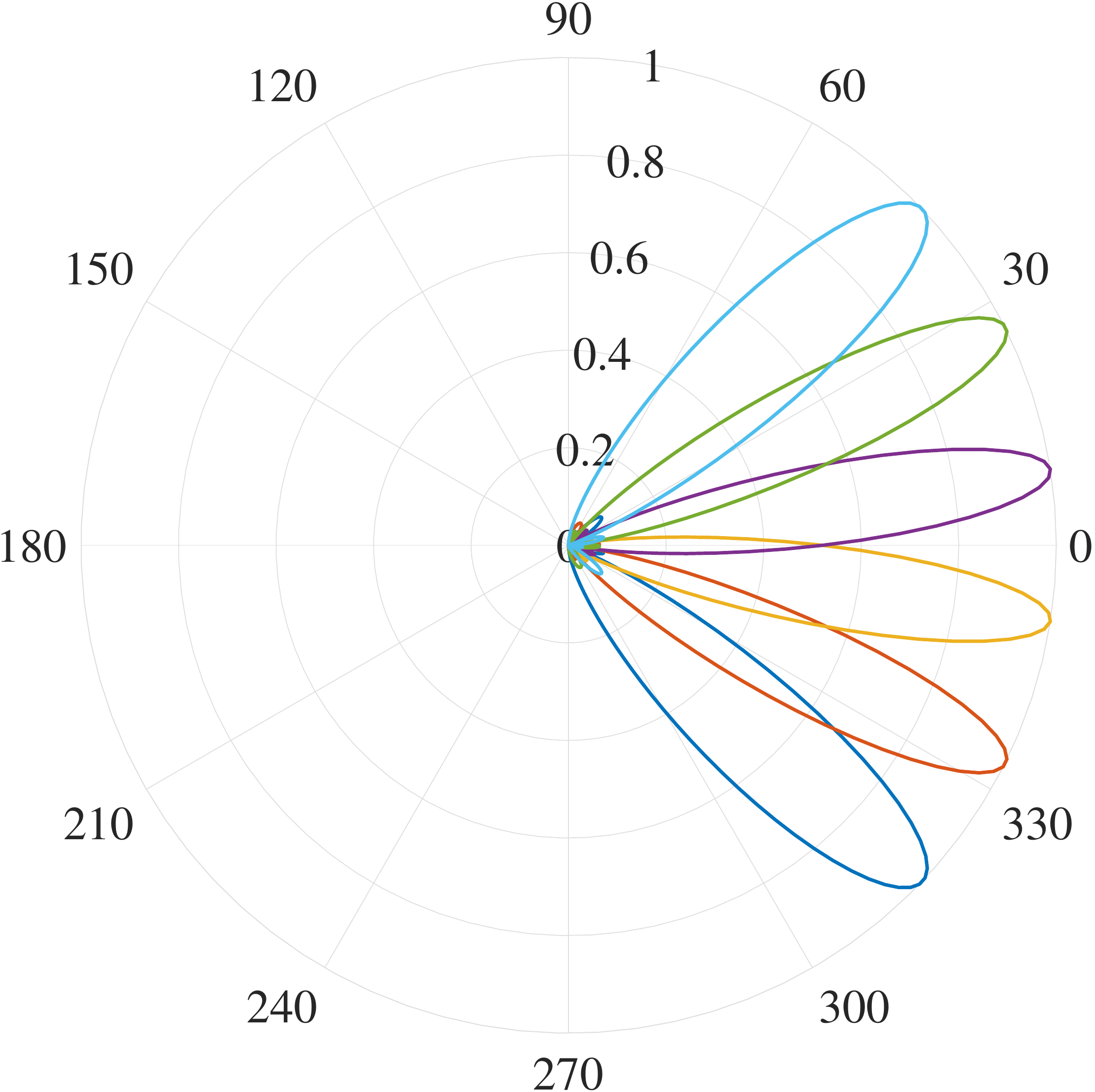}
\caption{Normalised radiation pattern of six  beams in a sector at $\Big[ \dfrac{-5\pi}{20},\dfrac{-3\pi}{20},\dfrac{-\pi}{20},\dfrac{\pi}{20},\dfrac{3\pi}{20},\dfrac{5\pi}{20}\Big]$ in azimuth plane }
\label{fig:BeamSectoral}
\end{figure} 
  The antenna array structure considered for generating beams for the BF-PDCCH is shown in Fig.~\ref{fig:AntennaStructure}. 
As per 3GPP specifications~\cite{1338901}, the rectangular panel array is described by the following tuple ($M_g,N_g,M,N,P$) , where,
$M_g$ and $N_g$ represent the number of panels in the vertical and horizontal direction, $M$ and $N$ represents the number of antenna elements with the same polarisation in the vertical and horizontal direction in each panel, and $P$ represents the panel is either single polarised ($P$=1) or cross polarised ($P$=2). The uniform rectangular array considered in this paper has the configuration of $ (1,1,4,8,2)$, where 4 elements are placed in vertical, 8 elements in horizontal and each being a cross-polarised has 64 antenna elements in total. With this given antenna structure, the effective array factor (AF) is calculated as follows.
  
\begin{multline}
%AF=\sum_{m=1}^{M}{e^{j(m-1)(kdz\sin\phi\cos\theta+\beta_z)}}\\ \sum_{n=1}^{N}{e^{j(n-1)(kdy\sin\phi\sin\theta+\beta_y)}} 
AF=\sum_{m=1}^{M}\sum_{n=1}^{N}{(e^{j(m-1)(kdz\cos\phi\sin\theta+\beta_z)})}\times \\ ({e^{j(n-1)(kdy\sin\phi\sin\theta+\beta_y)})} 
\label{eqn:RadPat}\end{multline}
where 
$k=\dfrac{2\pi}{\lambda}$, $dy$ and $dz$ are the antenna element spacing in horizontal and vertical, $\phi$ and $\theta$ are the azimuth and the elevation angles and, $\beta_y$ and  $\beta_z$ are the phase excitations for the antenna elements in the y and z axes respectively.  Note that as shown in the Fig.~\ref{fig:AntennaStructure}, $\phi$ and $\theta$ are the angles with respect to x-axis and z-axis respectively. 

%The beams can be generated in the desired direction by configuring the phase shifters $\beta_y$ and $\beta_z$. The phase shifters are configured such that the array factor is maximum in the desired direction. The maximum of AF from equation (\ref{eqn:RadPat}) is achieved by equating the exponential powers to zero.
%\begin{align*}
%\\
%kdy\sin\phi_0\sin\theta_0+\beta_y=0 &\implies \beta_y=-kdy\sin\phi_0\sin\theta_0 \\
%kdz\sin\phi_0\cos\theta_0+\beta_z=0 &\implies \beta_z=-kdz\sin\phi_0\cos\theta_0
%\end{align*}

\begin{figure}
\raggedleft
\includegraphics[width=10.2cm ]{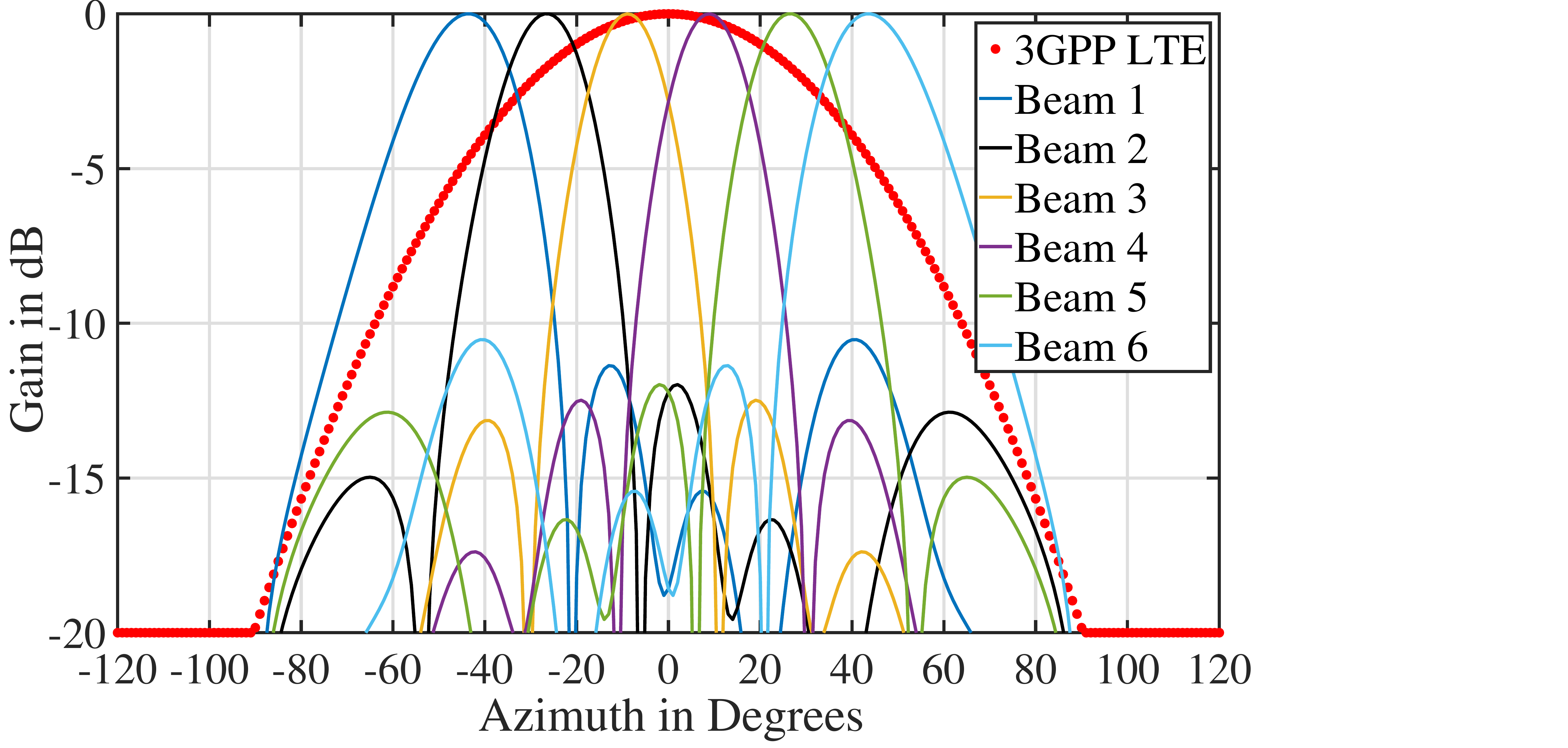}
\caption{Antenna pattern of six beams in the proposed BF-PDCCH compared against that of 3GPP LTE}
\label{fig:BeamLogy}
\end{figure}
%Thus $\beta_y$ and $\beta_z$ are configured based on the desired direction of the beam ($\phi_0,\theta_0$).
 A sample set of six beams covering the $120^{\circ}$ sector is generated according to the procedure mentioned above and is presented in the Fig.~\ref{fig:BeamSectoral}. The desired directions of the beams are assumed to be at six positions equi-distant from one other in the region of  [$-60^\circ , 60^\circ$] in the azimuth plane. Further, for each beam, the gains in the azimuth plane are plotted in the Fig.~\ref{fig:BeamLogy} and are compared against the 3GPP sectoral pattern. The sectoral pattern of legacy LTE is generated as per the 3GPP specifications~\cite{25.996}. Note that the side lobes are below the 10 dB level for each beam and all the six beams cover the entire $120^\circ$ sector similar to the legacy LTE sectoral pattern. 

For the transmission of data in this desired beams, the resource elements are multiplied with respective beam weights. A resource element is beamformed or transmitted in a particular direction by precoding it with the required beam pattern as follows.

\begin{equation}
\overline x^{i}_{k,l}=w_i.x_{k,l}
\end{equation}
\begin{multline}
%w_i=\sum_{m=1}^{M}{e^{j(m-1)(kdz\sin\phi\cos\theta+\beta_z(i))}}  \\\sum_{n=1}^{N}{e^{j(n-1)(kdy\sin\phi\sin\theta+\beta_y(i))}} 
w_i=\sum_{m=1}^{M}\sum_{n=1}^{N}({e^{j(m-1)(kdz\cos\phi\sin\theta+\beta_z(i))}})  \times \\({e^{j(n-1)(kdy\sin\phi\sin\theta+\beta_y(i))})} 
\label{eqn:beamweights} 
\end{multline}
 where, $x_{k,l}$ and $\overline x^{i}_{k,l}$ are the resource elements at $(k,l)$ index of frequency-time before and after beamforming respectively, $w_i$ has the beam weights for $i^{th}$ beam, and $\beta_y(i)$ and $\beta_z(i)$ are  the phase shifters generated for the  $i^{th}$ beam. Next, we present the BF-PDCCH design.
\section{Proposed Beamformed PDCCH design}
In this section, we initially explain the constraints for designing beamformed PDCCH as per 3GPP specifications. Then, we propose a novel beamformed PDCCH design and in the end, we discuss and analyse the performance of the proposed scheme.
\label{sec:BF-PDCCH}
%\subsection{3GPP Constraints}

In multi-user MIMO, initially, the best beams for each user are identified (a clear explanation of identifying the best beam for a user is provided in Section~\ref{sec:FlowChart}). These beams are spatially well separated, and thus, the transmission is done simultaneously in each beam with minimal interference. This way, multi-fold improvement is achieved in the network throughput. We consider $P$ beams are active in the sector all the time to implement multi-user MIMO.  The primary synchronisation signal (PSS), the secondary synchronisation signal (SSS), PBCH and PDCCH assume same channel characteristics as they rely on CRS. Hence, all of these have to behave similarly in terms of spatial configuration~\cite[Section 6.8.4]{1536211}. Thus, if PDCCH has to be beamformed, then it forces CRS, PBCH and PSS/SSS also to be beamformed.  With all these constraints a PDCCH beamforming has to be designed which can improve the PDCCH capacity.
\subsection{Proposed Scheme}
With all the constraints mentioned earlier, we propose the search space design for beamformed PDCCH as follows. Consider $P$ beams active in the sector all the time. All the common signals PSS, SSS, PBCH, PCFICH, PHICH and CRS are transmitted in all the beams. Thus there is no impact on CRS channel estimation based reception for common channels as both the CRS and the common channels observe a similar channel (a detailed mathematical explanation for the same is presented in the Section~\ref{sec:LLS} and is shown with system level simulations in the Section~\ref{sec:SLS}). 

PDCCH has two search spaces, CSS and USS. Since CSS has to address all the users present in the sector, CSS has to be present in all the beams exactly at same CCEs. In order to schedule USS differently in each beam, a mechanism is needed to identify the best beam for each user. For this purpose, we depend on the sounding reference signal transmitted in the uplink from the user. The implementation procedure for finding the best beams is presented in Section~\ref{sec:FlowChart}. Now with the best beam assigned to each user, the USS is scheduled differently in each  beam. Further, some users are at the cell edge/beam edge and experience a large interference with this dynamic scheduling of different data streams in each beam. Hence, the USS beamforming is applied only for the users who are at the boresight of the beam. Therefore, USS is split into two search spaces, USS~-~1 and USS~-~2. The search space for BF-PDCCH is thus divided into three regions:
\begin{figure}
\centering
\includegraphics[width=9cm]{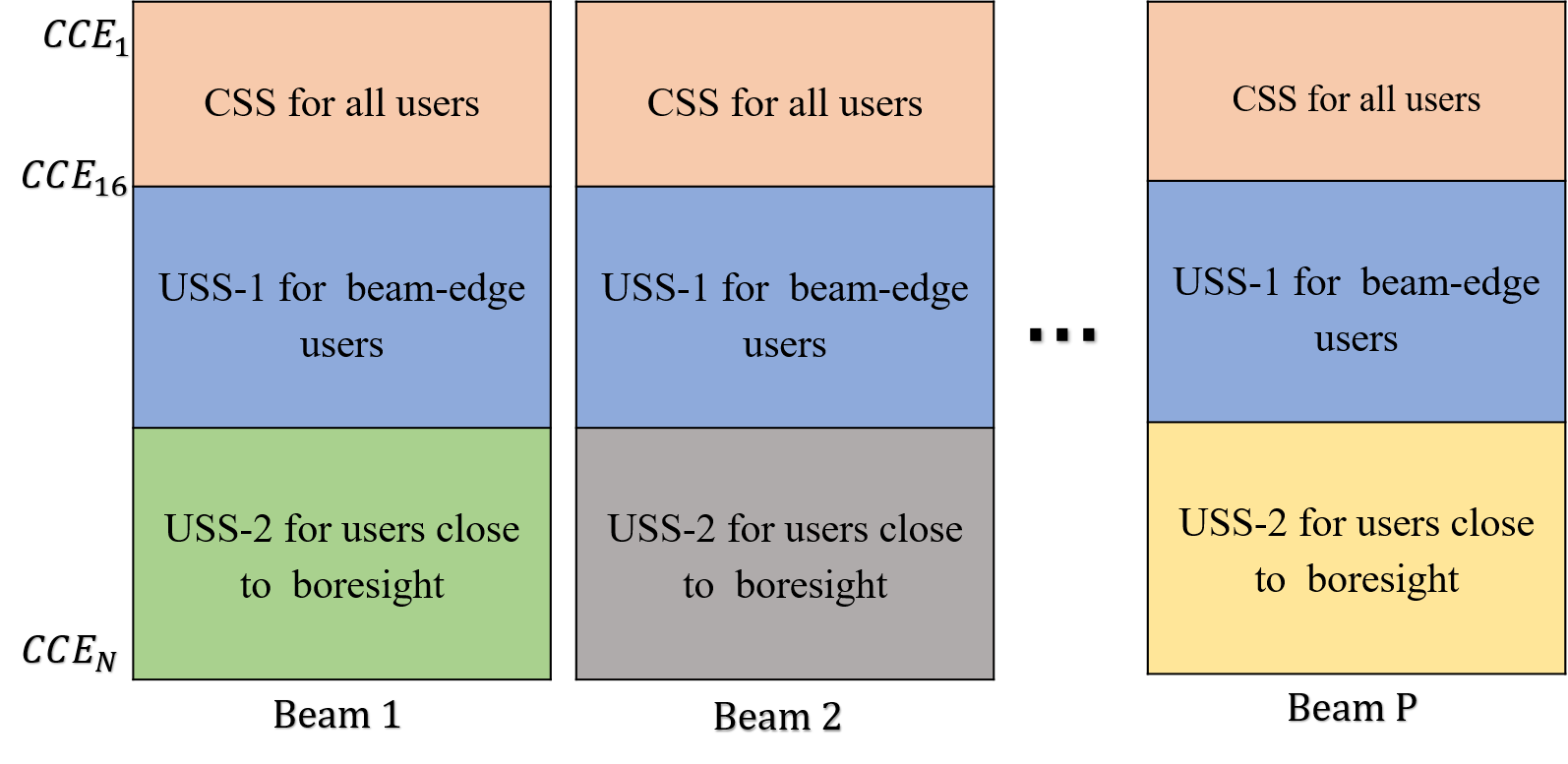}
\caption{ Search space design of proposed  BF-PDCCH}
\label{fig:BFPDCCH}
\end{figure} 
\begin{itemize}
\item Irrespective of the available bandwidth CSS is present in the first 16 CCEs of the PDCCH as in current 3GPP specifications~\cite{1536213}. Since CSS is common to all the users in the sector, it is present in all the beams.
\item  USS~-~1 has DCIs addressing the users at cell edge/beam-edge and hence, is present in all the   beams.
\item USS~-~2 has the DCIs addressing the users at boresight. Different DCIs are scheduled in each  beam and hence, the improvement in PDCCH capacity is achieved by USS~-~2.
\end{itemize}
An illustration of the proposed BF-PDCCH design is presented in Fig.~\ref{fig:BFPDCCH}.

Note that the users are not aware of PSS/SSS, common channels and the PDCCH being beamformed. They decode all the channels as per the 3GPP specifications. The users are scheduled in USS~-~2 only when the base station receives the sounding reference signal from the user. Until then the user will be scheduled in USS~-~1.  Further,  the CSS and the USS regions do not deviate from the current 3GPP specifications. There is no extra signalling for the user to indicate where it has to perform the blind decoding.   Only the base station has the notion of USS~-~1 and USS~-~2, and it intelligently schedules users in USS~-~1 and USS~-~2. The user is not aware of the PDCCH being beamformed. It considers the region of USS~-~1 and USS~-~2 as a regular USS in the legacy LTE PDCCH and does the blind decoding in a similar fashion as it does in the legacy LTE PDCCH. Next, we present the impact on the link level performance with the implementation of the proposed BF-PDCCH scheme.

\subsection{Performance Analysis}
\label{sec:LLS}
In this section, we present the impact on the performance of various channels because of the proposed beamforming design.
\subsubsection{\textbf{Channel estimation of common channels, CSS and USS-1}}

 \begin{figure*}
\begin{subfigure}{.5\textwidth}
\includegraphics[width=9.8cm]{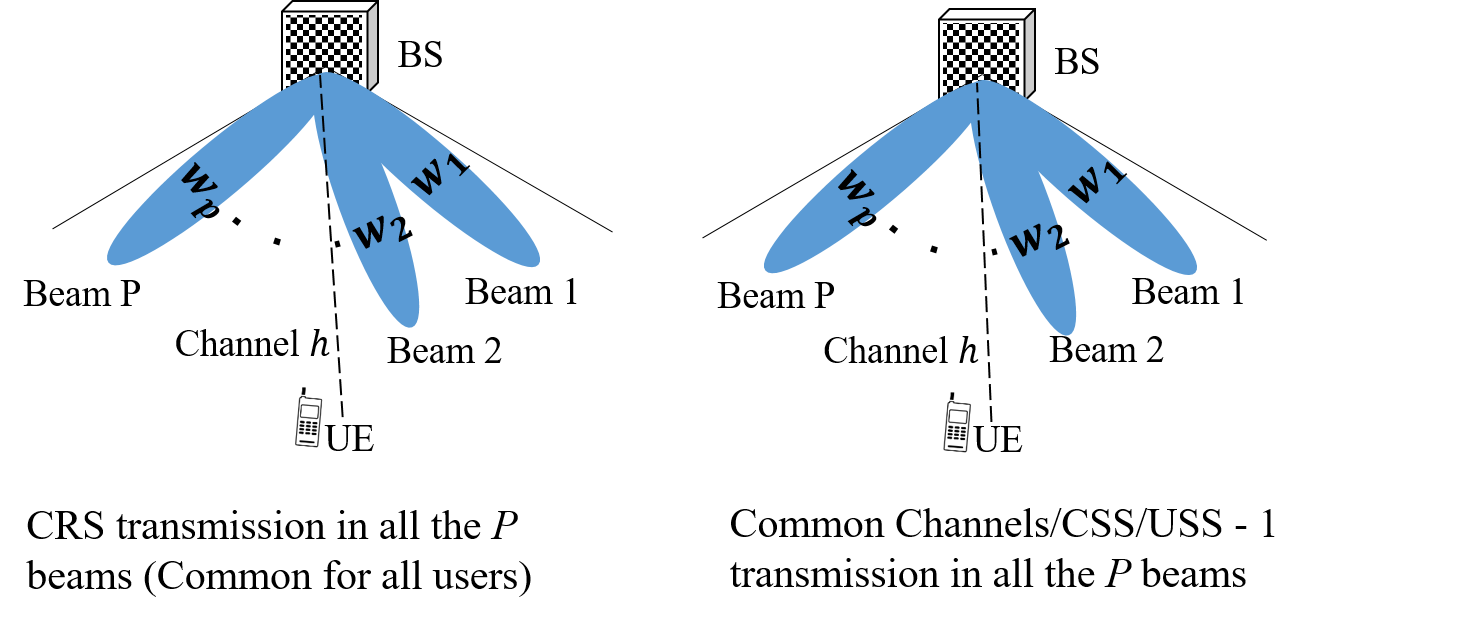}
\caption{CRS and Common channels/CSS/USS~-~1 transmission}
\label{fig:CSS}
\end{subfigure}
\hspace{0.cm}
\begin{subfigure}{.5\textwidth}
\includegraphics[width=8.2cm]{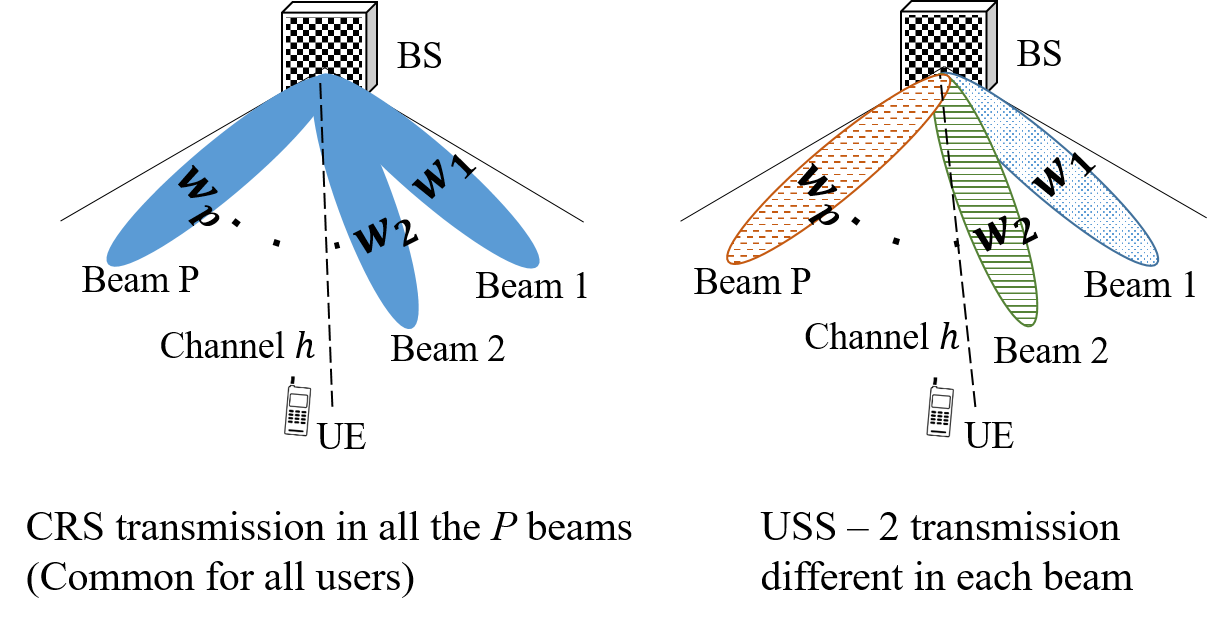}
\caption{CRS and USS~-~2 transmission }
\label{fig:USS}
\end{subfigure}
\caption{CRS, USS~-~1 and USS~-~2 transmission with the proposed BF-PDCCH}
\end{figure*}

Since CRS and USS~-~1 are present in all the  beams, the users observe the same channel for both of them as shown in the Fig.~\ref{fig:CSS}. %The same is formulated below in (\ref{eqn:CSS1})-(\ref{eqn:CSS2}).
 Let $y_{crs}$ and $y_{u_1}$ denote the received CRS and the USS~-~1 symbols, $x_{crs}$ and $x_{u_1}$ denote the transmitted CRS and USS~-~1 symbols, $w_i$ denote the beam weights as per (\ref{eqn:beamweights}), $h$ and $n$ denote the observed channel and the noise respectively. The estimated channel $h_{estimate}$ is formulated as follows.

\begin{align}
\nonumber
&y_{crs}&&=\sum_{i=1}^{P} h w_ix_{crs}+n\\
&\nonumber 
 h_{experienced}&&=\sum_{i=1}^P h w_i\\
&h_{estimate}&&=\dfrac{y_{crs}}{x_{crs}}=\sum_{i=1}^P hw_i +\hat{n}=h_{experienced}+\hat{n}
\label{eqn:CSS1}
\end{align}

\begin{align}
\nonumber
&y_{u_1}&&=\sum_{i=1}^{P} h w_ix_{u_1}+n\\
\nonumber
&y_{equalized}&&=\dfrac{y_{u_1}}{h_{estimate}}\\
&&&=\hat{x}_{u_1}+\tilde{n}
\label{eqn:CSS2}
\end{align}
% where $y_{crs}$ and $y_{data}$ are the received CRS and the data symbols, $x_{crs}$ and $x_{data}$ are the transmitted CRS and the data symbols, $w_i$ has the beam weights as per (\ref{eqn:beamweights}), $h$ is the channel and $n$ is the noise observed.

Since $x_{crs}$ and $x_{u_1}$ are transmitted in all the beams $i=1,\ldots,P$ and both observe the same channel $\sum_{i=1}^P hw_i$, the decoding has minimal impact when the  $h_{estimate}$ is used to equalise the received data $y_{u_1}$.

\subsubsection{\textbf{Channel estimation for the USS~-~2}}

In USS~-~2, different data is present in each beam as shown in the Fig.~\ref{fig:USS}. For a user with data transmitted in $i^{th}$ beam, the received CRS ($y_{crs}$), the received data ($y_{u_2}$), the estimated channel ($h_{estimate}$) and the equalized data ($y_{equalized}$) are formulated in \eqref{eqn:USS1}-\eqref{eqn:USS2}

\begin{align}
 \nonumber
& y_{crs}&&=hw_ix_{crs}+\sum_{j=1,j\neq i}^{P} hw_j x_{crs}+n \\
&\nonumber h_{experienced}&&=hw_i\\
&\nonumber h_{estimate}&&=\dfrac{y_{crs}}{x_{crs}}\\
&\nonumber &&=hw_i+\sum_{j=1,j\neq i}^P hw_j +\hat{n}
&\\&&&=h_{experienced}+h_{interference}+\hat{n}
\label{eqn:USS1}
\end{align}
 
\begin{align} 
\raggedleft
 \nonumber
 &y_{u_2}&&=hw_ix_{u_2}^i+\sum_{j=1,j\neq i}^{P} hw_j x_{u_2}^j+n&\\
&\nonumber y_{equalized}&&=\dfrac{y_{u_2}}{h_{estimate}}&\\
&&&= \hat{x}_{u_2}^i+x_{interference}+\tilde{n}&
\label{eqn:USS2}
\end{align}
% where $\alpha_j$ is the inter beam leak from the $j^{th}$ beam.

 Note that for the user in $i^{th}$ beam the channel experienced by the data is $hw_i$. Since same $x_{crs}$ is present in all the   beams, the user estimated channel would be $\sum_{i=1}^Phw_i$. 
Thus, the estimated channel has errors caused by the other active beams. Further, in addition to this incorrect channel estimation error,  there is an extra interference from other beams on the data as well. Both of these errors are  represented as $h_{interference}$ and $x_{interference}$ in \eqref{eqn:USS1}-\eqref{eqn:USS2}. Thus, there is an impact on the decoding of data for the USS~-~2.  Hence, with this BF-PDCCH, the operating signal to interference plus noise ratio (SINR) of the user will drop, and the user requires a comparatively large AL when transmitted in USS~-~2.   To solve this issue, we increase the AL for the user and allocate the user with more resource elements. However, we can compensate for this increased AL by packing more users within the existing resources by using the beam-specific scheduling for BF-PDCCH, i.e., multi-user MIMO concepts on the PDCCH. Note that with the proposed BF-PDCCH, the increase in the channel capacity can be achieved only through USS~-~2.
 Next, we present the procedures for the implementation of the proposed scheme.

\section{Procedures for the implementation and evaluation of BF-PDCCH}
\label{sec:implementation}

%\section{Simulation Results and Discussion}
A system level simulation typically abstracts the link layer characteristics. For this, a block error rate (BLER) vs. signal to interference plus noise ratio (SINR) curve is generated for different modulation and coding schemes which is thereafter used in a system level simulation.  For the proposed BF-PDCCH evaluation, we need to further abstract the channel estimation errors.  Hence, we initially present the link level simulations and conclude on the abstraction to be used in the system level simulations. Then, we present the implementation procedure for the proposed scheme.
%Then we present system level simulations for current 3GPP PDCCH, EPDCCH and the proposed BF-PDCCH schemes. 
\subsection{Link Level Simulations}
\subsubsection{\textbf{Abstraction of channel estimation errors for USS~-~2}}
\begin{figure}[t!]
\centering
\includegraphics[width=3.5 in, height=2.0 in]{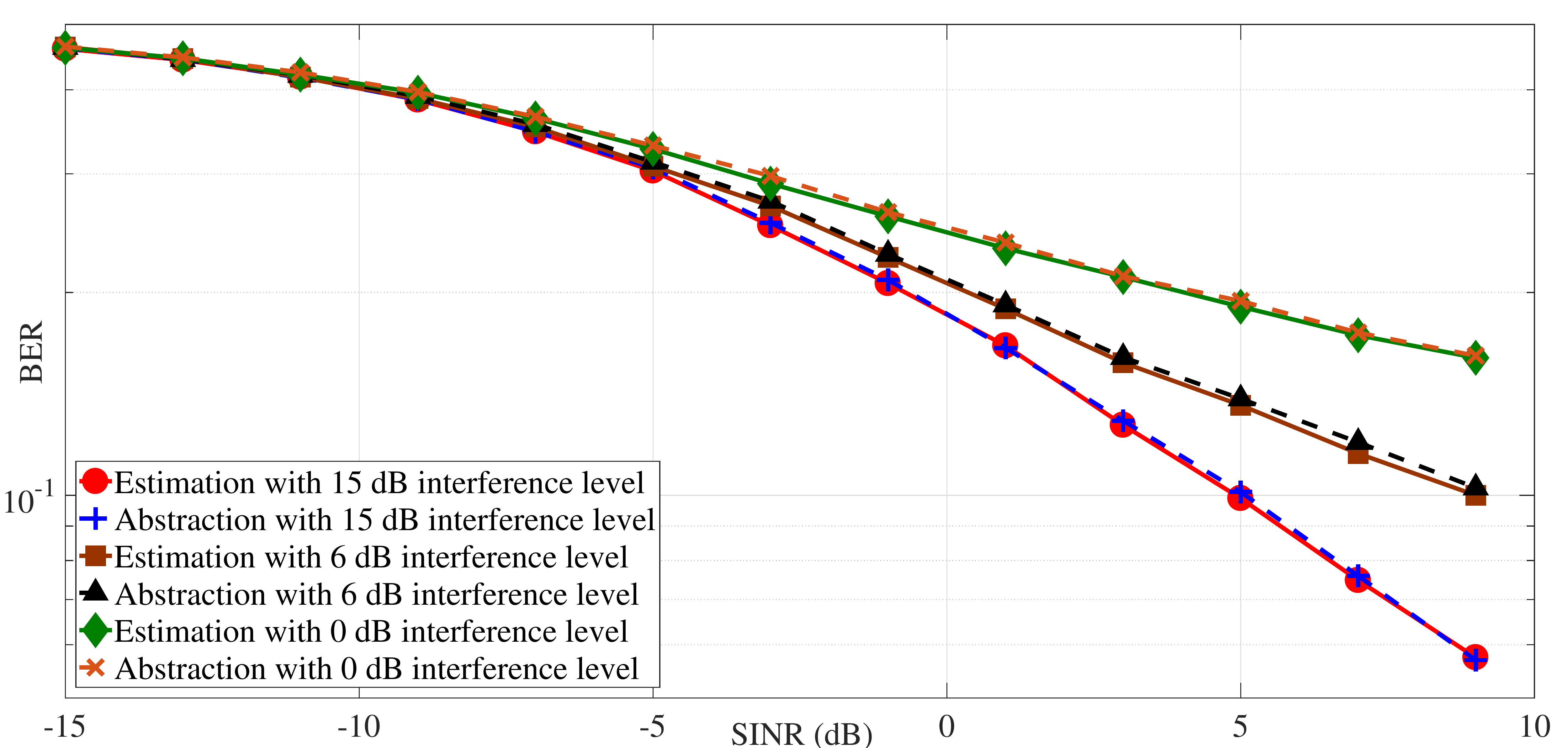}
\caption{ BER for both the estimation and abstraction for various interference levels}
\label{fig:abs}
\end{figure}

From (\ref{eqn:USS1})-(\ref{eqn:USS2}),  it can be observed that the degradation in the SINR of a user is mainly because of the interference observed in the channel estimation and the data.   Let $\alpha_j$ denote the inter-beam leak from $j^{th}$ beam of the current sector.  As shown in \eqref{eqn:USS1}-\eqref{eqn:USS2}, there is an impact of interference from other beams while estimating the channel $h_{estimate}$ and similar interference levels are seen while equalising the data $y_{u_2}$. Hence, a twice the interference from the other  beams ($2 \times \alpha_j$) in the current sector is added as an additional interference to compensate the both  effects and the SINR is formulated as shown in (\ref{eqn:abs}).   
\begin{equation}
SINR_{abs}=\dfrac{\Vert hw_i\Vert^2}{{\sigma_n^2}+2\sum_{j=1,j\neq i}^P\alpha_j}
\label{eqn:abs}
\end{equation}
%where $\alpha_j$ is the inter beam leak from the $j^{th} $ beam of the current sector.%and $\delta$ is the pilot gain (the number of pilots averaged in the link level simulations).

We have performed link level simulations with channel estimation and the abstraction mentioned above. The simulations are carried for Rayleigh fading channel, QPSK modulation and various inter beam leak ($\alpha_j$) levels. For the estimation curves, the channel is estimated in the presence of interference from the other beams as per (\ref{eqn:USS1})-(\ref{eqn:USS2}), and the bit error rate (BER) is calculated. Whereas for the abstraction, the impact of the channel estimation errors is captured in the SINR as per (\ref{eqn:abs}) and the BER is calculated.   The BER results for both the estimation and the abstraction are presented in the Fig.~\ref{fig:abs}. It can be observed that, the abstraction of the channel estimation errors gives similar performance as that of the estimated ones and hence, can be used for the system level simulations.

\subsubsection{\textbf{Abstraction of the block error rate (BLER) for mapping SINR to AL  }}

\begin{table}
\centering
\caption{Link level simulation parameters }
\setlength\extrarowheight{2pt}
\begin{tabular}{|m{4cm}|m{3.5cm}|}
\hline
\textbf{Parameter} & \textbf{Value}\\ \hline
DCI Size (format 1) & 31 bits\\ \hline 
CRC length & 16 bits\\ \hline
Channel Coding&tail biting convolution code\\ \hline
Modulation & QPSK\\ \hline
Bandwidth & 10 MHz\\ \hline
Channel Model & AWGN \\ \hline
Aggregation levels & \{1, 2, 4, 8\}\\ \hline
Antenna configuration (BS, UE)& 1, 2 \\\hline
\end{tabular}
\label{tab:link}
\end{table}

\begin{figure}[t!]
\includegraphics[width=3.5 in,height=2.1in]{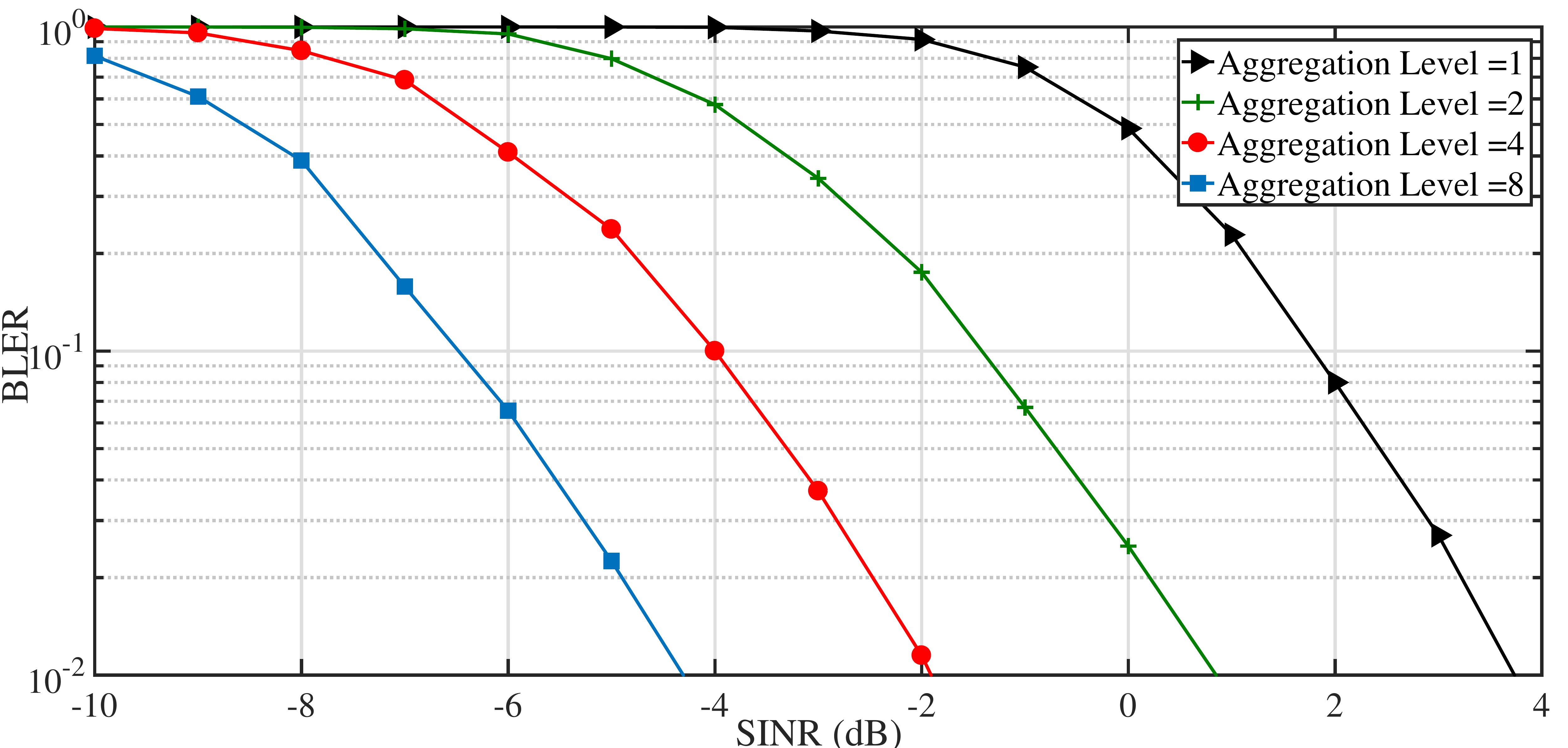}
\caption{BLER curves for various aggregation levels}
\label{fig:BLERabs}
\end{figure}

The link level simulations are carried out for the parameters mentioned in the Table~\ref{tab:link} and the block error rate curves (BLER) are plotted in Fig.~\ref{fig:BLERabs} for various ALs.  The results presented in the Fig.~\ref{fig:BLERabs} are aligned with \cite{nokia} and the 3GPP specification~\cite{1536101}. For the link abstraction in system level simulations, we consider 1\% block error rate for target SINR~\cite{nokia}  and perform the mapping of SINR to allocate an AL for each user. This a standard and a widely used mechanism for mapping the rate to SINR~\cite{nokia} and hence, we use the same in this paper. For a given SINR of each user, a minimum AL is chosen such that the BLER is less than 1\% at that SINR. 

The above two abstractions are used in the system level simulations while evaluating the performance of the BF-PDCCH and various other 3GPP mechanisms. Next, we present the implementation algorithm for the proposed scheme.

\subsection{Implementation algorithm for the proposed scheme}
\label{sec:FlowChart}
\begin{figure}
\centering
\includegraphics[width=9cm]{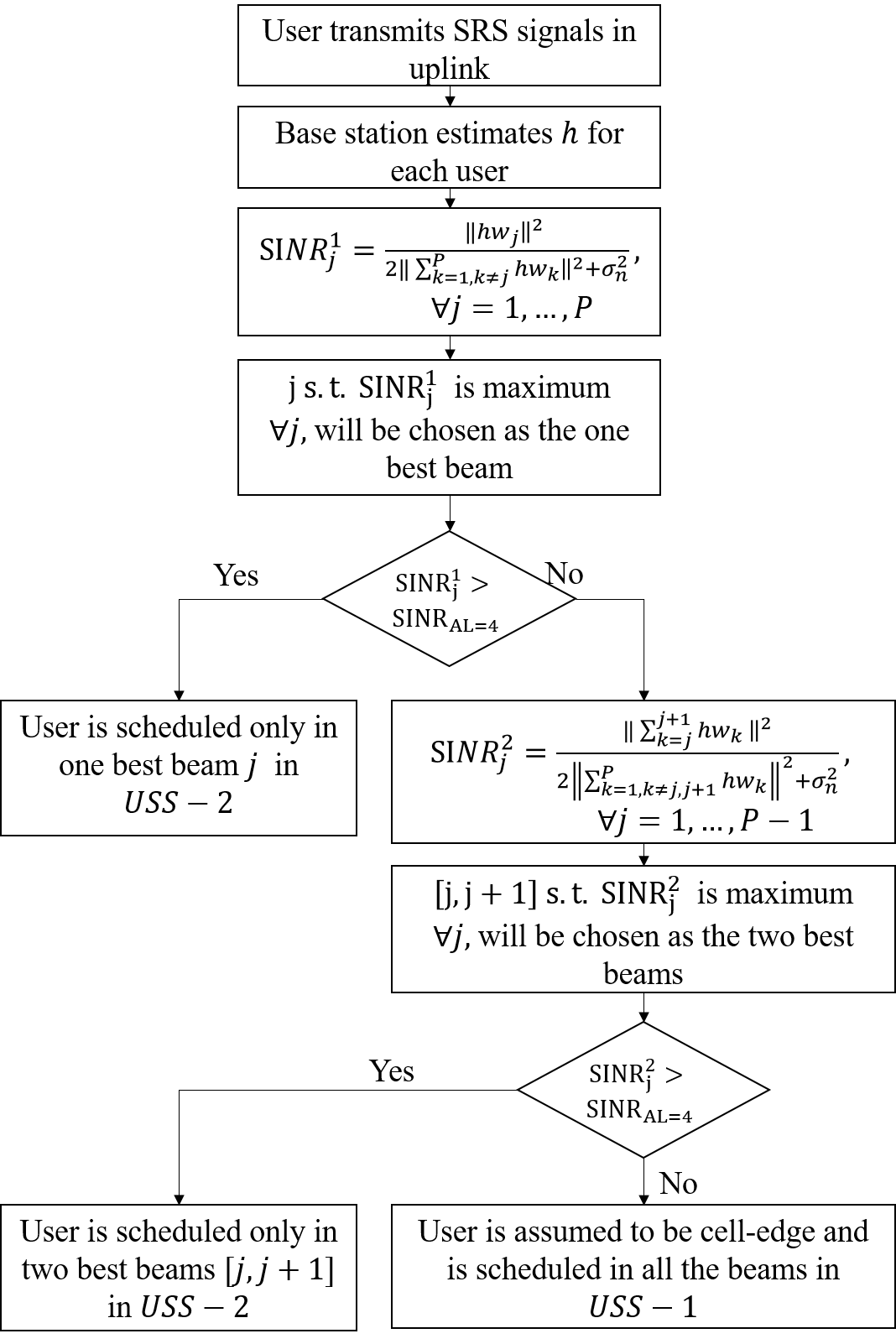}
\caption{ Flow chart depicting the implementation of the proposed scheme}
\label{fig:FlowChart}
\end{figure} 

\begin{figure}
\centering
\includegraphics[width=7.5cm]{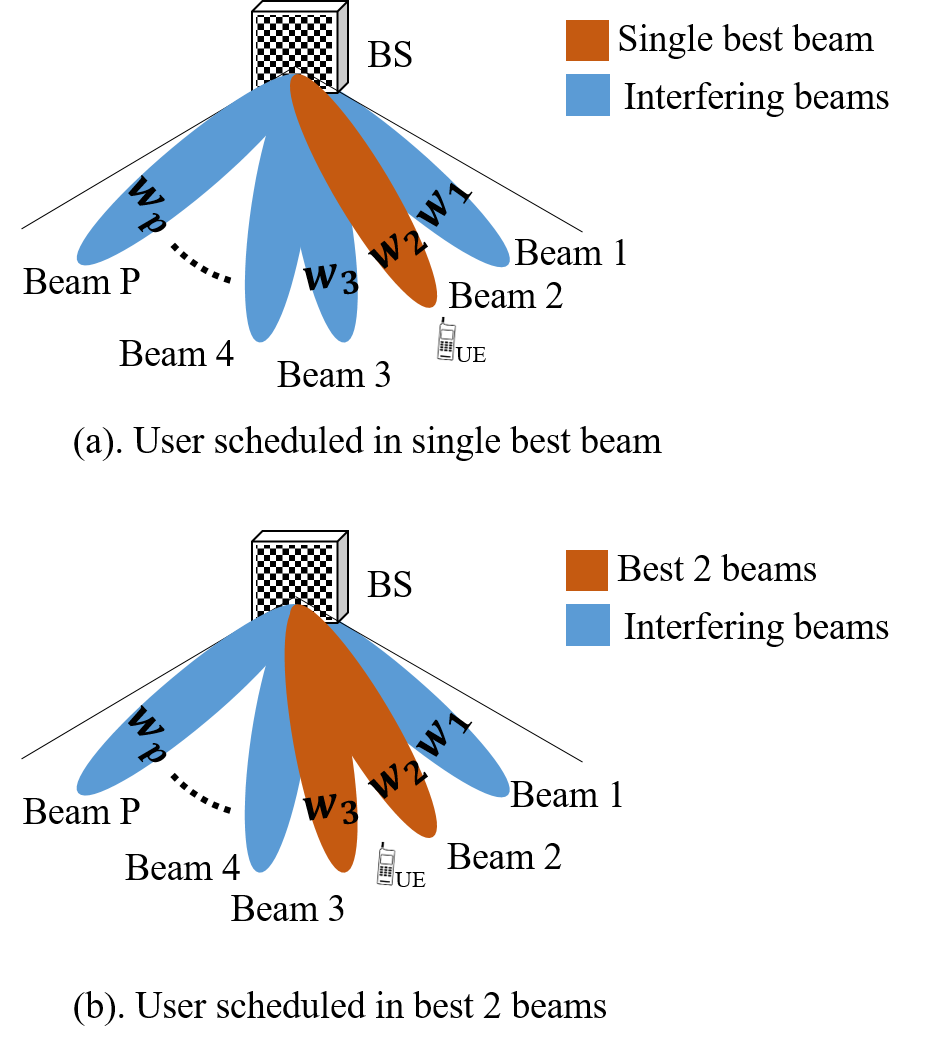}
\caption{ Pictorial representation of user scheduling in 1 and 2 best beams respectively}
\label{fig:BeamFit}
\end{figure}
The user transmits sounding reference signals (SRS) in the uplink. The base station receives the channel coefficients ($h$) on all the antenna from each user. The beam weights ($w_i$) for each specific beam are calculated as mentioned in \eqref{eqn:beamweights}. The $SINR_j^1$ is defined as the signal to interference-plus-noise ratio when the data is transmitted for a user in $j^{th}$ beam and interference is observed from all the other beams. According to \eqref{eqn:abs}, $SINR_j^1$ is calculated  as follows.
\begin{equation}
SINR_j^1= \dfrac{\left\Vert hw_j\right\Vert^2}{2\left\Vert\sum_{k=1, k\neq j}^{P}hw_k\right\Vert^2+{\sigma_n^2}}, \forall j=1,\ldots,P
\label{eqn:SINR1}
\end{equation}
The $j$ for which $SINR_j^1$ is maximum is chosen as the best beam for the user. For this chosen $j$, $SINR_j^1$ is compared against the minimum SINR required for $AL=4$ in legacy LTE case. If $SINR_j^1>SINR_{AL=4}$, then user is scheduled only in $j^{th}$ beam.  The reason for this is explained with an example as follows. Consider the following worst case scenario where eight users require AL=1 in legacy LTE PDCCH scenario, and the same eight users need AL=4 in BF-PDCCH scenario with eight beams ($P=8$). When scheduled in legacy LTE PDCCH, the users need 8 CCEs. But in the case of BF-PDCCH, even though they require 4 CCEs each when scheduled in a single best beam, they will be spatially multiplexed across the 8 beams in the same 4 CCEs and will need just 4 CCEs in total. Note that if the users require AL=1, 2 with BF-PDCCH then the multiplexing gain increases further, and when the users require AL=8 with BF-PDCCH, the gain completely disappears.  Hence, with eight beams active in the sector, we consider the users with a maximum of AL=4 for USS~-~2. 
%This is because, in a worst case scenario even if the user takes $AL=4$ with $SINR_j^1$, we could schedule $P$ such users in the other active beams. Thus, when $P =8$, eight users can be scheduled in just four CCEs. 

If the $SINR_j^1<SINR_{AL=4}$, then the user is checked for two best beams in a similar fashion as follows.
 \begin{equation}
SINR_j^2= \dfrac{\left\Vert\sum_{k=j}^{j+1}hw_k\right\Vert^2}{2\left\Vert\sum_{k=1, k\neq [j,j+1]}^{P}hw_k\right\Vert^2+{\sigma_n^2}}, \forall j=1,\ldots,P-1
\label{eqn:SINR2}
\end{equation}

$SINR_j^2$ is defined as the signal to interference-plus-noise ratio when data is transmitted in two best beams and interference is observed from the other beams. A similar procedure like earlier is followed and best beams $[j, j+1]$ are chosen such that $SINR_j^2$ is maximum. The user is then scheduled in those two best beams if  $SINR_j^2>SINR_{AL=4}$. Note that both the above transmissions will be scheduled in USS~-~2. If the $SINR_{j}^2\leq SINR_{AL=4}$, then the user is assumed to be in cell-edge/beam-edge and the data is transmitted for the user in all the  beams in USS-1. All the above procedure is presented as a flow chart in Fig.~\ref{fig:FlowChart}, and a pictorial representation of user scheduling is shown in Fig.~\ref{fig:BeamFit}. Next, we present the simulation results.

\section{Simulation Results and Discussion}
%\subsection{System Level Simulations}
\label{sec:SLS}
%The parameters for the system level simulations are considered as per the 3GPP specifications~\cite{1336873} and are presented in Table~\ref{tab:Sim}. The evaluation for various schemes is performed as follows.  The users are dropped randomly in a cell site. For the given parameters in the Table~\ref{tab:Sim}, the SINR is calculated for each user.  Based on the BLER abstraction mentioned earlier, we map the aggregation levels to each user for this SINR. For a given bandwidth and number of OFDM symbols, the available CCEs vary. We assume half the available CCEs carry downlink DCIs. Within this limited number of CCEs, we schedule the users with different ALs. 
%For the EPDCCH case, we consider an extra four PRBs of resources available from the data channel.  For the BF-PDCCH case, we consider four active beams as mentioned in the Table~\ref{tab:Sim} and use the abstraction of channel estimation errors explained earlier while calculating the SINR of any user.%\subsection{PDCCH Evaluation}
%\subsection{EPDCCH Evaluation}
%\subsection{BF-PDCCH Evaluation}
\begin{table}[t]
\caption{System level simulation parameters}
\setlength\extrarowheight{2pt}
\begin{tabular}{|m{3.5cm}|m{4cm}|}
\hline
\textbf{Parameter} & \textbf{Value}\\ \hline
Cell layout & 7 cell sites, 3 sectors/site\\ \hline 
Inter-site distance & 500 m\\ \hline
BS antenna height & 25m\\ \hline
UE antenna height & 1.5m\\ \hline
Carrier frequency & 2.4 GHz\\ \hline
BS transmit power & 44 dBm\\ \hline
Number of antennae (BS, UE)& 64, 2 (BF-PDCCH),\newline 1, 2 (LTE PDCCH)\\\hline
Bandwidth & 20 MHz\\ \hline
Channel model & 3D UMa in TR 36.873~\cite{1336873}\\ \hline
Direction of selective beams  & 
%\begin{itemize}
%\item 
Azimuth:[$\dfrac{-3\pi}{16},\dfrac{-\pi}{16},\dfrac{\pi}{16},\dfrac{3\pi}{16}$]\ %\item
Elevation:[$\dfrac{9\pi}{16},\dfrac{11\pi}{16}$]
%\end{itemize}
\\ \hline
Available DL CCEs & 42 (with CFI=3 OFDM symbols)\\ \hline
Aggregation levels & \{1, 2, 4, 8\}\\ \hline
Extra PRBs for EPDCCH & 4 PRBs\\ \hline

BS antenna element radiation pattern & According to TR 36.873~\cite{1336873}\\ \hline
\end{tabular}
\label{tab:Sim}
\end{table}

The parameters for the system level simulations are considered as per the 3GPP specifications~\cite{1336873} and are presented in Table~\ref{tab:Sim}. The evaluation for various schemes is performed as follows.  The users are dropped randomly in a cell site. For the given parameters in the Table~\ref{tab:Sim}, the SINR is calculated for each user.  While calculating the SINR, the interference from other sectors ($I_{wrap}$) is modelled using wrap around algorithm mentioned in~\cite{wrap}.   When the data is transmitted in one best beam, the user observes interference from all the remaining $P-1$ beams. Similarly, when the data is transmitted in two adjacent beams, the user observes interference from all the remaining $P-2$ beams.  Apart from the interference from the other sectors and other  beams, the user observes errors in the channel estimation because of CRS re-use happening across the beams. These channel estimation errors are accounted in SINR calculation as per the abstraction procedure mentioned in \eqref{eqn:abs}-\eqref{eqn:SINR2}. Thus, the SINR for various configurations is calculated as follows. 
  
%While calculating SINR, the wrap-around model is enabled and interference is considered from all the other sectors in a ring. Considering the interference from other sectors and channel abstraction as per \eqref{eqn:SINR1}-\eqref{eqn:SINR2}, the SINR for various configurations is calculated as follows.
\begin{itemize}
\item SINR with data in one best beam:
\begin{align*}
&\widehat{SINR}_j^1=&&\dfrac{\left\Vert hw_j\right\Vert^2}{I_{wrap}+2\left\Vert\sum_{k=1, k\neq j}^{P}hw_k\right\Vert^2+{\sigma_n^2}}, \\&&&\hspace{3cm}\forall j=1,\ldots,P\\\\
\end{align*}
\item SINR with data in two best beams:
\begin{align*}
&\widehat{SINR}_j^2=&&\dfrac{\left\Vert\sum_{k=j}^{j+1}hw_k\right\Vert^2}{I_{wrap} +2\left\Vert\sum_{k=1, k\neq [j,j+1]}^{P}hw_k\right\Vert^2+{\sigma_n^2}},
\\& &&\hspace{3cm}\forall j=1,\ldots,P-1\\
\end{align*}
\item SINR with data in all the  beams:
\begin{align*}
&\widehat{SINR}^P=\dfrac{\left\Vert\sum_{k=1}^{P}hw_k\right\Vert^2}{I_{wrap} +{\sigma_n^2}}
\end{align*}
\end{itemize}

\begin{figure}[t!]
\centering
\includegraphics[width=3.5 in, height=2.2 in]{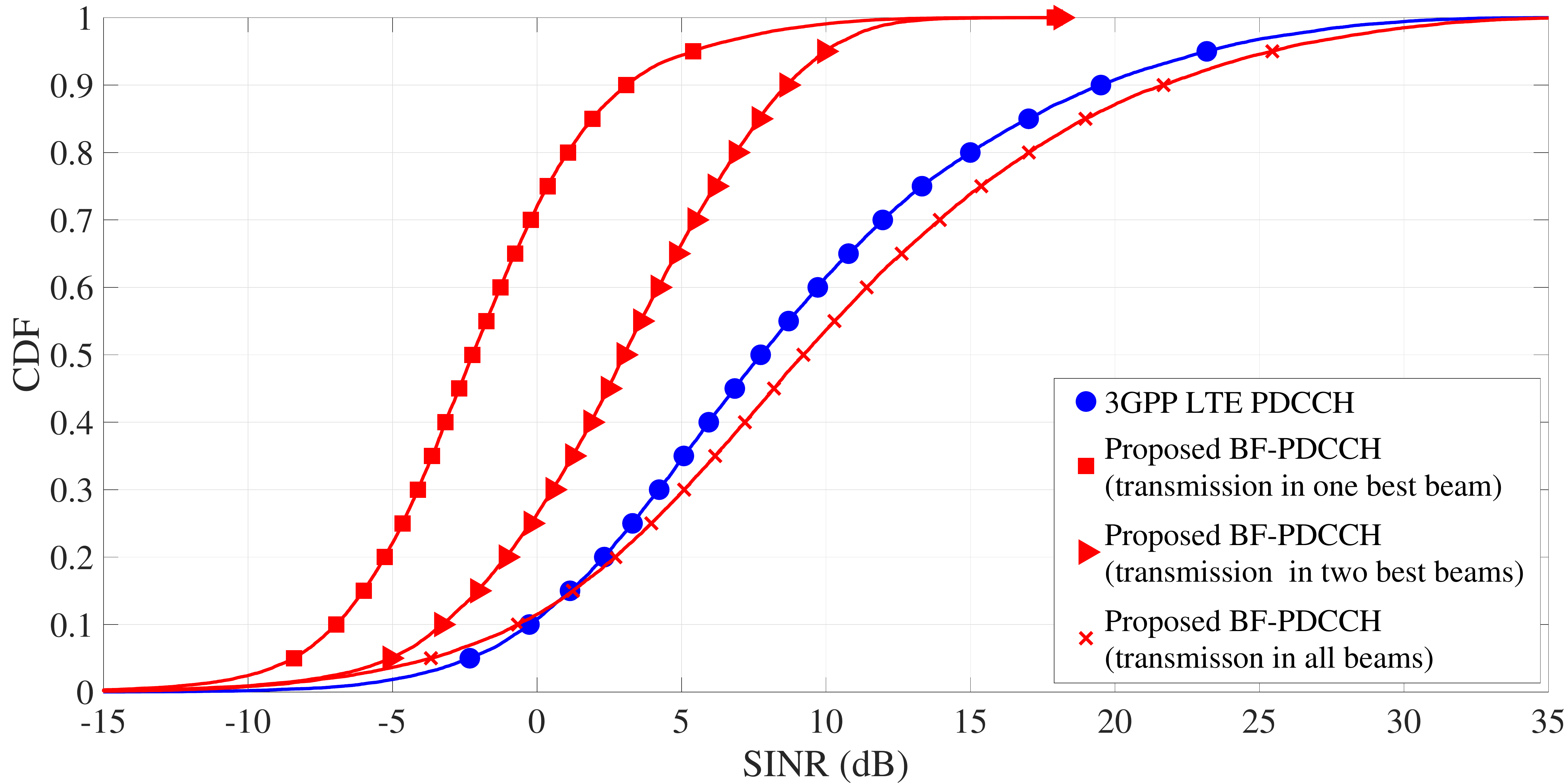}
\caption{ Variation of SINR of users with proposed and the existing 3GPP schemes}
\label{fig:SINRcdf}
\end{figure} 

 In Fig.~\ref{fig:SINRcdf}, the SINR distribution of users with the various configurations of the active number of beams is presented and compared against that of the legacy LTE. The SINR curve for 3GPP LTE PDCCH is generated for a 1 tx and 2 rx antennae case. The cumulative distribution function (CDF) of the SINR for the 3GPP LTE PDCCH in Fig.~\ref{fig:SINRcdf} is aligned with the 3GPP calibrations mentioned in \cite{36.814}.    The proposed BF-PDCCH with data transmitted in all eight  beams has the mean SINR value greater than LTE PDCCH.
  With eight beams active all the time, each beam gets $\dfrac{1}{8}^{th}$ of the total transmit power. However, we see an improvement in the mean SINR value because of beamforming done using the  64 antenna elements. Note that, when compared to the legacy LTE case, there are null regions between the beams, and hence, the users observing the lower SINRs has also increased.   
  
  Further, the SINR distribution has been analysed for the case where the data is transmitted in one best beam, and best two adjacent beams.  From Fig.~\ref{fig:SINRcdf}, it can be observed that more than 30\% of the users have an SINR range above 0~dB and hence, require AL=2 even when the data is transmitted in one beam. Thus for all these users, if MU-MIMO is enabled in other beams, there will be an eight-fold improvement in the network capacity. Note that when the data is transmitted in the best two adjacent beams, more than 70\% of the users have an SINR range above 0 dB. Considering all the above facts, we perform the scheduling for BF-PDCCH according to the algorithm mentioned in Section~\ref{sec:FlowChart} as follows. Initially, based on the SRS feedback, one best beam is identified for the user. The AL required for the base station to transmit a DCI to the user in  that best beam is calculated. Based on the BLER abstraction mentioned in the earlier section, the SINR is mapped to the required AL. If the AL is less than eight with data in one best beam, it helps in achieving improvement in the network capacity. This is because, if there are eight other users with AL=4 which have other beams as their best beams, they can be scheduled across different beams. The base station can thus schedule eight users in only 4 CCEs.  When the user requires AL=8 with the transmission in one beam, the AL required while transmitting in two adjacent beams is checked and a similar procedure as earlier is followed. If the user needs AL=8 in both the scenarios, then the user is assumed to be a cell-edge user and DCI will be scheduled for the user in all the active eight beams.  The scheduling procedure is continued for all the users until the CCEs in a TTI are exhausted. The remaining users are scheduled in the successive TTIs in a similar manner.

\begin{figure}
\centering
\includegraphics[width=3.5 in, height=2.2 in]{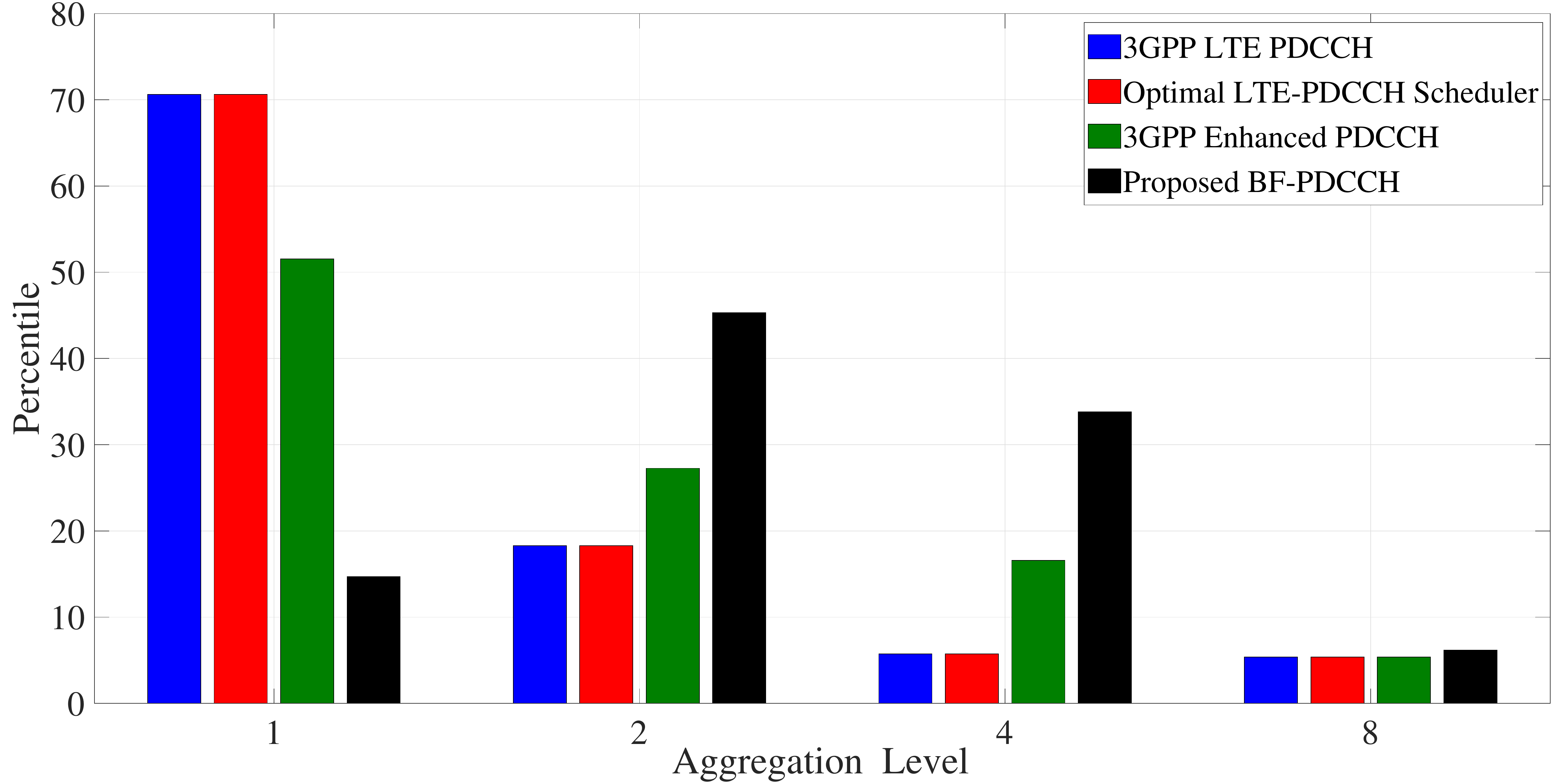}
\caption{ Histogram of average number of users scheduled across various ALs}
\label{fig:ALbar}
\end{figure}

\begin{figure}[t!]
\centering
\includegraphics[width=3.5 in, height=2.2 in]{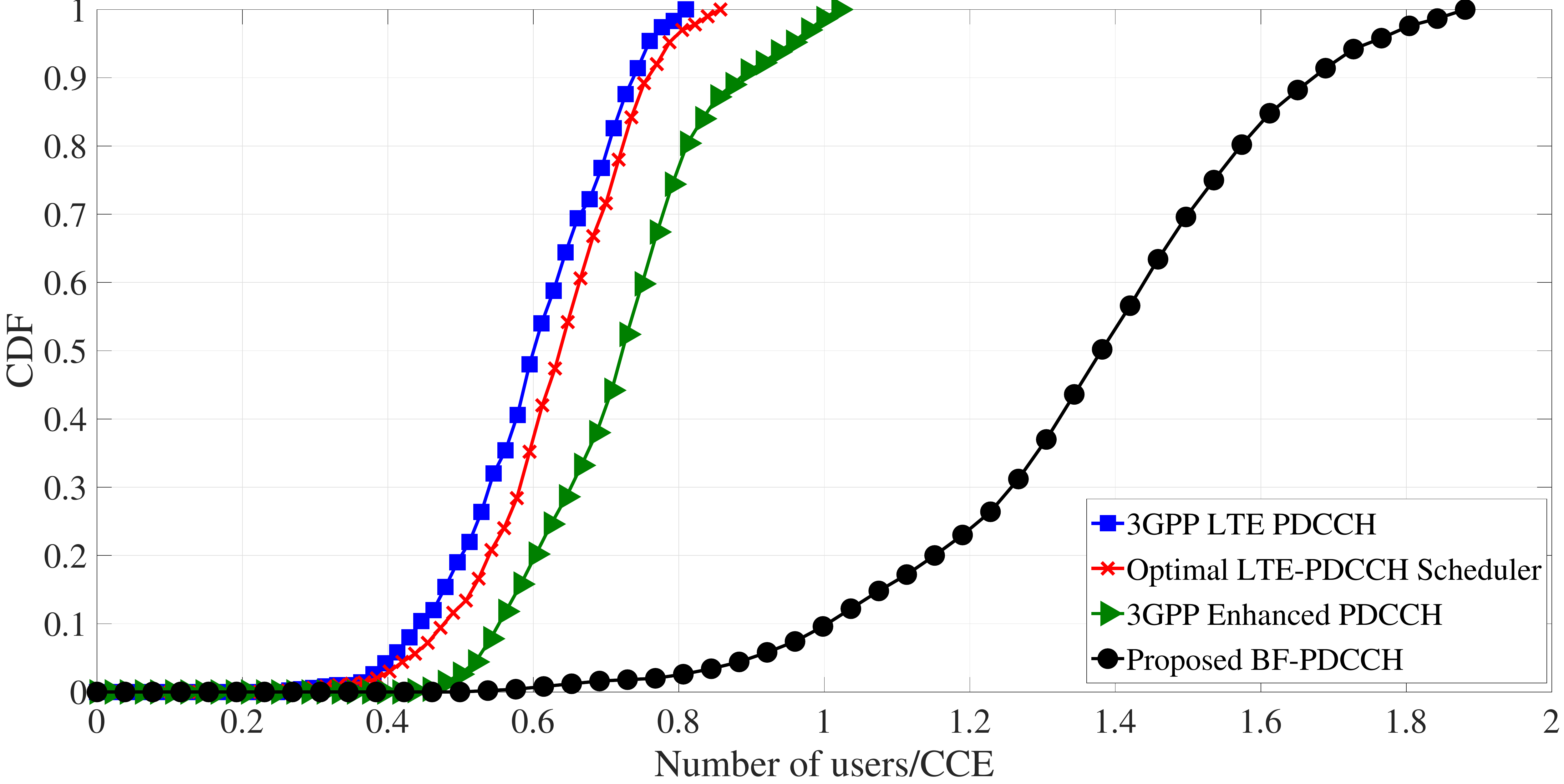}
\caption{ CDF plot of number of users scheduled/CCE}
\label{fig:UEcdf}
\end{figure}

  For a given bandwidth and number of OFDM symbols, the available CCEs vary. We have performed the simulation with 500 users dropped in a sector and for a bandwidth of 20 MHz with PDCCH active in 3 symbols in each TTI.  For 20 MHz bandwidth and  PDCCH in 3 symbols, the number of CCEs available is 84 as per 3GPP specifications~\cite{1536211}. We assume that half the CCEs are available for both downlink and uplink grants. Hence, in the simulation, the users are scheduled in the available 42 downlink CCEs.
A similar procedure is followed for PDCCH and EPDCCH. For the EPDCCH case, the additional PRBs could be 2, 4 and 8~\cite{1536213}. In our simulations, we consider 4 PRBs of resources from the data channel are available for EPDCCH. Further, we have implemented the optimal scheduling procedure presented in~\cite{Balamurali} where the users are scheduled based on a set packing algorithm to increase the control channel capacity. We have compared its performance against the proposed BF-PDCCH design in Fig.~\ref{fig:ALbar},~\ref{fig:UEcdf}.

In Fig.~\ref{fig:ALbar}, the allocation of the aggregation level with the proposed BF-PDCCH is compared against that of 3GPP LTE PDCCH, optimal LTE-PDCCH scheduler~\cite{Balamurali} and enhanced PDCCH. The optimal LTE-PDCCH scheduler presented in~\cite{Balamurali} schedules the users in the available control channel region by minimising the resource wastage. It has similar AL allocation as of the 3GPP LTE PDCCH.
Compared to 3GPP LTE PDCCH, enhanced PDCCH has lesser users with AL=1 and more users with AL=2, 4.  When a user is scheduled in enhanced PDCCH region, the probability of having AL=1 is reduced because of the MU-MIMO scheduling and thus, more number of users are scheduled with AL=2, 4.  A similar change is observed with the proposed BF-PDCCH as well. However, since the entire control channel is beamformed for MU-MIMO, the probability of users with AL=1 is further less than that of the enhanced PDCCH case and thus, the probability of users with AL=2, 4 has increased. As shown in Fig.~\ref{fig:SINRcdf}, when compared with 3GPP LTE PDCCH, there are more users with lower SINR values even when the data is transmitted in all the beams.  Thus, the percentile of users with AL=8 is more than that of the 3GPP LTE PDCCH case. Note that even though the AL is increased, the network capacity will be enhanced because of the MU-MIMO feature.

\begin{table}
\centering
\setlength\extrarowheight{3pt}
\caption{Average users scheduled in a TTI with the proposed and current 3GPP mechanisms}
\label{tab:Res}\begin{tabular}{|m{5cm}|m{2.8cm}|}
%{|m{2.8cm}|m{2.8cm}|m{1.5cm}|m{1.5cm}|}
%\hline
%Scheme&Legacy LTE PDCCH& Enhanced PDCCH&BF-PDCCH\\\hline
%Mean users/TTI&18.04&24.01&37.55\\\hline
\hline
 \textbf{Scheme} &\textbf{Average users/TTI}\\\hline
 3GPP LTE PDCCH& 25.1\\\hline
 Optimal LTE-PDCCH Scheduler~\cite{Balamurali} &26.4\\\hline
 Enhanced PDCCH& 36.0\\\hline
 Proposed BF-PDCCH & 58.3\\\hline
 
\end{tabular}
\end{table}

In Fig.~\ref{fig:UEcdf}, the CDF of the number of users scheduled per CCE is presented. The number of CCEs available is more in enhanced PDCCH because of the extra four PRB resources, and hence, for a fair comparison, the normalisation is carried out with the available CCEs for each case.  
The 3GPP LTE PDCCH has the least performance among all the schemes.  The optimal LTE-PDCCH scheduler~\cite{Balamurali} minimises the control channel resource wastage and hence, can accommodate more users in the control channel region. Thus, optimal LTE-PDCCH scheduler has comparatively more number of users scheduled per CCE. The EPDCCH has a better performance compared to the 3GPP LTE PDCCH and optimal LTE-PDCCH scheduler. However, note that EPDCCH uses an additional four PRBs from the PDSCH resources and also requires additional control signalling.  The proposed BF-PDCCH has better performance than all these schemes. The average number of users scheduled in a TTI are presented in the Table~\ref{tab:Res}. In the proposed BF-PDCCH scheme, even though the users experience a poor operating SINR and require large AL, because of the multi-user MIMO more number of users can be scheduled. From Fig.~\ref{fig:UEcdf}, it can be observed that more than two-fold improvement in the network capacity is achieved with the proposed BF-PDCCH.

\section{Conclusion and Future Work}

We proposed a novel beamforming design for the control channel of LTE which is aligned to the current 3GPP specifications and requires no changes at the user end.  Unlike the current 3GPP mechanisms of enhancing the capacity, the proposed scheme does not use additional resources from the data channel. We efficiently use the large antenna structure available at the base station and schedule more users in the PDCCH with FD-MIMO. For this, we rely on the sounding reference signals transmitted in the uplink to decide the best beam for a user. We then ingeniously schedule the users and enhance the PDCCH capacity.  We performed link level simulations and provided mechanisms to abstract the channel estimation errors caused by the proposed scheme. We then used these abstractions in the system level simulations to evaluate the performance of the proposed scheme.  We have shown that the proposed design always performs better than the state of art algorithms and the existing 3GPP schemes. In future, we will validate the performance of the proposed BF-PDCCH design by implementing it on the hardware test-beds.
\label{sec:CFW}
\bibliographystyle{./IEEEtran}
\bibliography{GW}
\end{document}